\documentclass[apjl,ip,twocolumn]{aastex631}
\usepackage{comment}
\usepackage{ifthen}
\usepackage[hang,tight]{subfigure}
\usepackage{nicefrac}

\usepackage{xcolor}

\newcommand{\rev}[1]{#1}

\newcommand{\forloop}[5][1]%
{%
\setcounter{#2}{#3}%
\ifthenelse{#4}%
	{%
	#5%
	\addtocounter{#2}{#1}%
	\forloop[#1]{#2}{\value{#2}}{#4}{#5}%
	}%
	{%
	}%
}%

\newcommand{\ctbd}[1]{}

\newcommand{\C}{\ensuremath{^{\circ}C\;}}

\newcommand{\sqarcsec}{\ensuremath{\Box^{\prime\prime}}}

\newcommand{\msqarcsec}{\ensuremath{\rm mag/\sqarcsec}}
\newcommand{\cdmsq}{\ensuremath{\rm cd\,m^{-2}}}

\newcommand{\tabr}[1]{\mbox{Table~\ref{tab:#1}}}

\newboolean{emulateapj}
\setboolean{emulateapj}{true}

\newboolean{astroph}
\setboolean{astroph}{true}

\shortauthors{Kocifaj et al.}
\shorttitle{Light Pollution by Proposed Reflect Orbital Space Mirrors}
\ifthenelse{\boolean{emulateapj}}{
    
}{
    
}

\begin{document}

\title{
Atmospheric Light Pollution by Proposed Reflect Orbital Space Mirrors
}

\correspondingauthor{G\'asp\'ar Bakos}
\email{gbakos@astro.princeton.edu}

\author[0000-0001-9277-4692]{Miroslav Kocifaj}
\affiliation{ICA, Slovak Academy of Sciences, Slovakia}
\affiliation{Faculty of Mathematics, Physics and Informatics, Comenius University in Bratislava, Slovakia}
\email{miroslav.kocifaj@savba.sk}

\author[0000-0001-7204-6727]{G\'asp\'ar \'A. Bakos}
\affiliation{Department of Astrophysical Sciences, Princeton University, NJ 08544}

\author[0000-0002-2095-734X]{Franti\v{s}ek Kundracik}
\affiliation{Faculty of Mathematics, Physics and Informatics, Comenius University in Bratislava, Slovakia}
\email{Frantisek.Kundracik@fmph.uniba.sk}


\begin{abstract}
\setcounter{footnote}{10}
ReflectOrbital is planning to launch a $18\times18$\,m mirror in space at a
$\sim$600\,km altitude to illuminate a $2.5$\,km radius circular patch on
Earth.  The pilot satellite, called EARENDIL-1, is a pathfinder for a
constellation of $\sim$50,000 larger mirrors, each with $54\times54$\,m in
diameter, with the explicit goal to provide space-borne illumination on
Earth.  We calculate the sky brightness caused by the Rayleigh scattering
and aerosol scattering of the incoming beam from one such satellite and by
the light reflected back by the illuminated patch, for various atmospheric
properties and ground albedos under cloud-free conditions.  We demonstrate
that the light pollution caused by one of these satellites is very
significant, altering the nighttime environment up to distances of
$\sim$30\,kms.  The results show that for an observer within the beam of a
single 54\,m satellite, the mirror will appear as a $-16.7$\,mag point-like
source, i.e.~about 4 magnitudes brighter than the full moon.  The diffuse
sky background will be similar to the dusk sky shortly after sunset; too
bright to see even the brightest stars.  From a distance of 14\,km, the glow
from a single mirror will exceed the luminance of the full moon sky for
majority of the sky.  From a distance of 34\,km the sky will still appear
brighter than the moon-lit sky in the direction of the beam.
%
%
If 400 such mirrors illuminate the same patch
simultaneously, the glow will be obvious from 80\,kms.
\setcounter{footnote}{0}
\end{abstract}


\section{Introduction}
\label{sec:intro}

The idea of lighting up Earth from space is not new.  The first documented
idea came from the pioneering theorist of astronautics, Hermann Oberth, in
the 1920s.  \rev{Oberth} described the concept of large orbital mirrors to reflect
sunlight onto Earth for illumination \citep{oberth1923,oberth1929}.  In the
1990s, the Russian Znamya program was a set of experiments investigating
deployable space mirrors for reflecting sunlight onto Earth
\citep{NASA1993Znamya,ESA_Znamya}.  In 1993, Znamya 2 successfully deployed
a $\sim$20\,m reflective membrane from the Mir space station, producing a
moving illuminated spot on Earth with the illumination level roughly
comparable to that of moonlight from the full moon.  A subsequent attempt
(Znamya 2.5) failed to deploy properly, underscoring the challenges
associated with large, lightweight orbital reflectors.

Recently, the startup company Reflect Orbital filed a request to the United
States Federal Communications Commission (FCC) under FCC File
No.~SAT-LOA-20230825-00198
\footnote{\url{https://licensing.fcc.gov/myibfs/displayLicense.do?filingKey=-487580}}
\citep{ROFCC:2025} to launch an $18\times18$\,m mirror in space as a
pathfinder experiment toward establishing a giant constellation of 50,000
mirrors, each with $54\times54$\,m in size.  \rev{The request was recently
approved by the FCC without considering the environmental and safety impacts. 
Plans from 
Reflect Orbital} lack the
quantification of the resulting, potentially significant light pollution in Earth's
atmosphere caused by such mirrors.  The light pollution comes from the
following sources: i) the scattering of the incoming, $\sim2.5$\,km radius
beam in the atmosphere (air molecules and aerosols), ii) the scattering of
the light that is reflected back by Earth into the atmosphere, and iii) the
Mie scattering of the light on clouds (with an estimated $\sim$67\% of the
Earth covered in clouds).  The plans also lack consideration of the
environmental damage caused by the light pollution due to the satellites.

In this paper, we describe the light pollution by performing a detailed
calculation of the scattered light in Earth's atmosphere using \rev{a}
3-dimensional Radiative Transfer Equation model, and we characterize the
night sky brightness distribution at varying distances from the center of
illuminated area\footnote{See \url{https://starryprinceton.org/scattering_calculations_renderings}
for visual renderings}.
We consider several combinations of the aerosol asymmetry
parameter (g) and the surface albedo ($\alpha$).  The calculations assume an
aerosol optical depth of $0.30$ at the reference wavelength of 550\,nm, a
value that is typical for many regions of the world
\citep{MODIS:2021,JASTP:2018,JGR:2004}.  For simplicity, we assume a
cloud-free atmosphere with the understanding that under cloudy conditions, 
e.g.~with a cloud deck at 10\,km altitude illuminated by the beam, the 
overall illumination distribution will be starkly different. 
Since the prototype
satellite is a pathfinder for the much larger mirrors, we performed
calculations for the large mirrors.  Scaling our results to the 18\,m
prototype mirror is trivial, as the light reflected by the 18\,m mirror is
1/9th that of the proposed 54\,m mirrors.

The paper is structured as follows.  In Sec.~\ref{sec:backofenvelope}, we
carry out a simple calculation of the light levels by the prototype (18\,m) and
large (54\,m) mirrors without any scattering.  For the rest of the paper we
only consider the 54\,m scenario.  In Sec.~\ref{sec:out} we
show the sky brightness for observers outside the illuminated area, while in
Sec.~\ref{sec:in} we carry out calculations for an observer standing in the
illuminated patch.  In Sec.~\ref{sec:moon} we compare the results to the sky
background due to the full moon, and in Sec.~\ref{sec:conc} we summarize our
conclusions.


\section{Ballpark Figures}
\label{sec:backofenvelope}

\begin{figure*}[!ht]
\centering
\includegraphics[width={0.5\linewidth}]{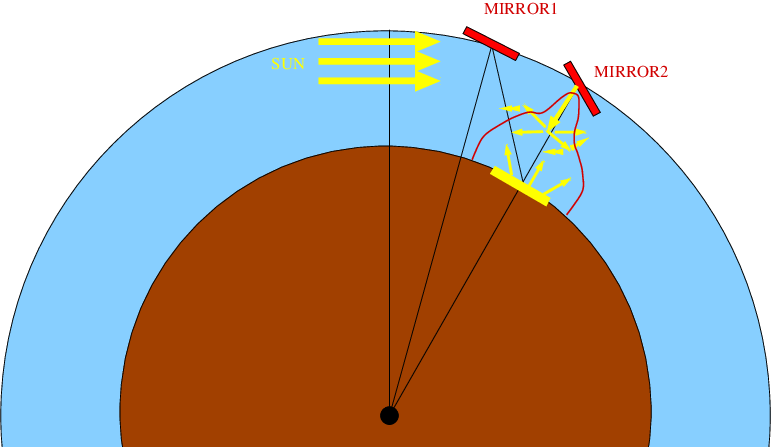}
\caption{
Schematic view of a proposed space-mirror illuminating the Earth.
The observer is past astronomical twilight, with the Sun 18\arcdeg\ below
the horizon. ``Mirror1'' is seen at a 45\arcdeg\ elevation from the
observer, while ``Mirror2'' is directly overhead. Yellow arrows indicate the
scattering of light in the atmosphere and reflected back from the ground. 
}
\label{fig:scheme}
\end{figure*}

First, we characterize the parameters of the 18\,m prototype (EARENDIL-1)
mirror.  For simplicity, we assume that the observer is standing right on
the terminator (right after sunset), with the mirror in zenith projecting
light directly underneath.  The mirror is tilted at $\epsilon = 45$ degrees
with respect to the direction of the Sun and also the direction of the
observer.  As we show later, different mutual positions of the observer
and the mirror within reasonable limits do not fundamentally change the
ballpark figures.  We assume a mirror size of $18 \times 18~\mathrm{m}$ and
a reflectance of 0.9, an altitude of the mirror of $H = 600~\mathrm{km}$, 
and an aerosol optical depth of $\tau_a =
0.3$, a Rayleigh optical depth of $\tau_R = 0.1$, leading to a combined
optical depth of $\tau = 0.4$ and an overall transmission of $T = e^{-\tau}
\approx 0.67$.  The solar irradiance is $F_\odot = 1361~\mathrm{W\,m^{-2}}$,
and the luminous efficacy is $110~\mathrm{lm/W}$ \citep{ies:2020}.
The projected dimensions of the mirror are:
$
18~\mathrm{m} \times (18 \cos 45\arcdeg)\,\mathrm{m} = 18~\mathrm{m} \times
12.7~\mathrm{m}, 
$
leading to an apparent angular size of 
$
\approx 6.2'' \times 4.4'',
$
and a projected
collecting area of $A_{\rm proj} = 18 \times 18 \times \cos45^\circ \mathrm{m^2} \approx
229~\mathrm{m^2}$.
To calculate the apparent brightness, we calculate the incident power using
the solar constant and the projected area as:
\[
P_{\rm in} = 1361 \mathrm{W/m^2} \times 229 \mathrm{m^2} \approx 3.12\times10^5~\mathrm{W} . 
\]
With the mirror reflectance of 0.9, the reflected power is:
\[
P_{\rm refl} = 0.9\,P_{\rm in} \approx 2.81\times10^5~\mathrm{W},
\]
and the transmitted power is:
\[
P_{\rm ground} = P_{\rm refl} e^{-0.4} \approx 1.88\times10^5~\mathrm{W}.
\]

Using the claimed spot area of 2.5\,km radius, and the area of the circle of
$
A_{\rm spot} = \pi (2500)^2 \mathrm{m^2} \approx 1.96\times10^7~\mathrm{m^2}, 
$
the irradiance is
\[
E = \frac{P_{\rm ground}}{A_{\rm spot}} = \frac{1.88\times10^5 \mathrm{W}}{1.96\times10^7\mathrm{m^2}} \approx 0.0096~\mathrm{W/m^2}
\]
and the illuminance is
\[
E_{\rm lux} \approx 0.0096 \mathrm{W/m^2}\times 110 \mathrm{lm/W} \approx 1.05~\mathrm{lux} . 
\]

To place this into context, the approximate horizontal illuminance due to
the full Moon near zenith is: $\sim 0.26~\mathrm{lux}$ \citep{Kyba:2017}, 
leading to a horizontal ground illuminance ratio of the prototype satellite to the full Moon of:
$1.05/0.26 \approx 4.0$.

The approximate apparent magnitude of the mirror, using the full Moon for
scaling, is:
\[
m = -12.74 - 2.5\log_{10}\left(\frac{1.05}{0.25}\right) \approx -14.3\,.
\]

Finally, with an orbital velocity of
$
v \approx 7.56~\mathrm{km/s}, 
$
the angular speed of the mirror is
\[
\omega = \frac{v}{H} \approx 0.72^\circ/\mathrm{s} \approx 2600~\mathrm{arcsec/s}. 
\]
In summary, the prototype mirror with $18\times18$\,m size will cause a
$\sim4$ times full-Moon illumination of $1.05~\mathrm{lux}$ over a $\sim
2.5~\mathrm{km}$ radius region, delivering $\sim 0.0096~\mathrm{W/m^2}$ of
energy, and will appear as a $-14.3$ magnitude point-like source, moving at
an approximate angular speed of $\omega \approx 0.72^\circ/\mathrm{s}$.

A similar calculation for the proposed, $54\times54$\,m square mirror leads
to a $\sim38$ times full-Moon illumination of $9.5~\mathrm{lux}$ (equivalent
to street-light level lighting) over the same $\sim 2.5~\mathrm{km}$ radius
region, delivering $\sim 0.086~\mathrm{W/m^2}$ of energy, and will appear as
a $-16.7$ magnitude, still point-like ($18.6'' \times 13.1''$) source.

If we now assume that the observer has just entered full darkness, i.e.,
past the astronomical twilight, with the sun elevation of $-18\arcdeg$, 
and we also assume that the mirror is at $45\arcdeg$
elevation (toward the West), simple geometry shows that the mirror is
$\alpha_M\approx13.25\arcdeg$ from the terminator as seen from the center of
the Earth, lagging behind the observer (who is already $\beta_O = 18\arcdeg$ past the
terminator; see Fig.~\ref{fig:scheme}).  The normal vector of the mirror is tilted by
$\epsilon\approx58.5\arcdeg$ with respect to the direction of the observer,
yielding a projected surface area of $\cos(58.5\arcdeg)\approx0.522$ that of
the entire mirror area.  Taking into account the slightly larger optical
depth due to the slant angle (factor of $e^{-0.414 \tau_0} =
e^{-0.414\cdot0.4} = 0.847$), using the same atmospheric parameters, assuming
that the size of the illuminated patch is still the same, we get that this
tilted mirror will be $\sim24$ times full-Moon illumination of
$5.95~\mathrm{lux}$, delivering $\approx 0.054~\mathrm{W/m^2}$ of energy,
and will appear as a $-16.19$ magnitude source.

In the same scenario, with the observer just entering full darkness ($\beta_O
= 18\arcdeg$), but with the mirror being in local zenith ($\alpha_M = 18\arcdeg$), 
the normal vector
of the mirror is tilted by $\epsilon=36\arcdeg$ with respect to the
direction of the observer, yielding a projected surface area of
$\cos(36\arcdeg)\approx0.809$ that of the entire mirror area.  The
atmospheric optical depth is as in the first scenario, when the observer was
straight on the terminator, because the mirror is still in zenith.  In this
situation the tilted mirror will be $\sim44$ times full-Moon illumination of
$10.9~\mathrm{lux}$, delivering $\approx 0.098~\mathrm{W/m^2}$ of energy,
and will appear as a $-16.85$ magnitude source.  In brief, the brightness of
the satellite will barely change between these configurations, and it is
reasonable to assume that the optimum satellite configurations would be used
for illumination, i.e., for an observer at $\beta_O$ degrees past sunset,
selecting the $\alpha_M$ longitude of the mirror such that illumination is
performed with a relatively large projected mirror size and at a small optical
depth.  In summary, our baseline assumption of the mirror being in zenith
and tilted at 45\arcdeg\ is good to about $\sim15-20$\%, and slightly
underestimates a situation whereby the observer is already past astronomical
twilight and the mirror is overhead.  The exact number will depend on
whether the mirrors are deformable, and the patch size is kept constant.

Using 400 mirrors, all at 45\arcdeg\ elevation, and assuming a mirror
reflectance of 0.9, the combined brightness will be $\sim2400$ lux, or about
$\sim10{,}000$ times brighter than the illuminance due to the full Moon, or
about $2.4\%$ of the illuminance due to the Sun.  These 400 mirrors can
deliver roughly $22~\mathrm{W/m^2}$ power, which is close to the lower limit
where solar panels can start operating.

\section{Sky Brightness Outside The Beam}
\label{sec:out}

We used the Modified Successive Orders of Scattering (MSOS) model
\citep{kocifaj:2018,kocifaj:2023}, which is a 3D Radiative Transfer Equation
model to compute the field of scattered radiation at any point in 3D space
(on a chosen 3D grid), both for downward‑ and upward‑propagating radiation. 
The original code was developed for light emitted upward from ground-based
sources and can have any angular emission pattern, such as a light cone.  If
the atmosphere was homogeneous throughout its entire volume, adapting the
MSOS code to the present case of an incoming beam from space would be
straightforward; we would simply compute the distribution of upward
radiation at the top of the atmosphere and then rotate the coordinate
system.  This would effectively move the ground‑based source to the top of
the atmosphere, and what we compute as upward radiation at the top of the
atmosphere would become downward radiation at the Earth’s surface.

However, the atmosphere is a vertically stratified medium: air and aerosol
densities decrease exponentially with altitude in a defined manner.  In our
calculations, we assumed a scale height for the air density of 8000\,m, and
a scale height for the aerosols of 2000\,m \citep{Qiu:2005}.  We kept the
MSOS model as it is, removed the ground‑based sources, and simulated the
nadir‑pointing mirror by defining a brightness field at the top of the
atmosphere (TOA) such that the radiance at zenith ($z = 0\arcdeg$) is set to
unity and is zero in all other directions.  We applied the same procedure
for all modeled altitudes above the Earth’s surface, with the brightness
reduced according to atmospheric extinction.  We then ran the computation
for 9 scattering orders to ensure reliable convergence of the model (in
principle, $\sim$3 orders are sufficient).

Because the MSOS model operates on a fixed spatial grid common to all
scattering orders, applying it directly to a narrow collimated beam presents
a resolution challenge.  A highly directional source concentrates the
optical signal within a small solid angle, meaning that at each altitude
level the beam footprint would be contained within a single voxel of the
standard grid.  Such a grid may be insufficient to resolve the generating
function of the first scattering order accurately (within the light cone). 

We therefore adopted a hybrid approach: the contribution of higher
scattering orders (as described in this section) were obtained from the MSOS
model, and the single-scattered radiance within the light cone was computed
analytically using the formalism described in Sec.~\ref{sec:in}, which
permits arbitrarily fine angular sampling at modest computational cost.
\rev{
Because the directional preference of the photons is progressively lost
with increasing scattering order, the higher-order radiance patterns become
gradually smoother and form a slowly varying continuum.  The fixed MSOS grid
resolves this continuum adequately.  Only the single-scattering (first-order)
field varies rapidly across the narrow cone, so the fixed grid cannot capture
its fine features.  Refining the grid to do so would impose the same fine
sampling on every order.  This would be costly in memory and computing time,
although only the first order requires it.  For this reason, the first-order
field within the cone is evaluated analytically, which permits arbitrarily
fine sampling.  The single-scattering and higher-order fields are then
superimposed to give the total radiance.  In practice this superposition is
realized through a pre-computed ratio.  From a full MSOS run, in which all
orders share the same grid, we form the ratio of the total (all-order)
radiance to the first-order radiance for each viewing direction.  The total
radiance is then recovered as the product of the accurate analytical
first-order field and this smoothly varying ratio.  The procedure retains the
full angular detail of the direct-beam scattering signal while capturing the
diffuse continuum produced by multiple scattering.  Since the total is a
single product of the accurate first-order field and a smooth ratio, the two
contributions neither overlap nor introduce a discontinuity.
}

The conversion to absolute units was done taking into account the mirror
area, its altitude above the Earth, the solar illuminance at the Sun–Earth
distance, and the solid angle subtended by the solar disk.  We assumed the
parameters listed in Sec.~\ref{sec:backofenvelope}.  In these calculations,
the direct beam from the mirror contributes only at the single sky position
corresponding to the mirror direction, and only for an observer located
directly within the illuminated patch.  At all other grid points, only the
diffuse radiation field is computed, which is the primary focus of our
calculations in this section.

The calculations include both scattered light and light reflected from the
Earth’s surface \citep[see Eq.~1]{kocifaj:2019}, where $\alpha$ is the
surface albedo.  We carried out calculations for albedos of $\alpha=0$ (zero
reflection), 0.2 (typical Earth) and 0.8 (fresh snow)
\citep{Sailor:2002,Liang:2001}.  The zero albedo calculations help in
establishing the fractional contribution of the light scattered by the
incoming beam, as compared to the light reflected back from the Earth in a
Lambertian manner.  We also used two different values of $g = 0.6$ and $g =
0.8$, where $g$ is asymmetry parameter in
scattering theory, i.e., the average cosine of the scattering angle over all
directions.  A value of $g=0.6$ is representative of fine-mode-dominated
aerosol (e.g., urban/continental conditions), while $g=0.8$ corresponds to
coarse-mode-dominated aerosol \citep{Kinne:2019} (e.g., maritime or
dust-laden atmosphere); together they bracket the range of asymmetry factors
reported for tropospheric aerosol \citep{Kinne:2013,Hatzianastassiou:2007}.

For each simulation we derive the map of the skyglow at various distances
from the ``epicenter'' of the beam, with all distances being outside the
beam.  (Calculations for the sky brightness for an observer inside the beam
are presented later in Sec.~\ref{sec:in}).  The values of the skyglow in
the figures is given in \cdmsq, and can be easily converted
\citep{Bara:2020} to magnitudes
per square arcsecond via the approximate formula of $
\mu\;[\msqarcsec] = 12.6 - 2.5\,\log_{10}\!\left(B\;[\cdmsq]\right)$.

\rev{
All calculations are performed at 550\,nm, a wavelength near the maximum
of the solar spectrum and near the maximum of the sensitivity of the human
eye, and it therefore dominates the radiance pattern.  Although blue light is
scattered more strongly, its share in the solar spectrum is small.  The
conversion from radiance to luminance is made through a constant of
proportionality (luminous flux per spectral radiant flux).  This uses the
fact that the relative angular distributions of sky radiance and luminance
under daylight conditions are known to be nearly identical
\citep{Torres:2008,Alshaibani:2020}, differing essentially only by such a
constant.  \citet{Igawa:2004} confirmed that the relative radiance and
luminance distributions of the sky can be described by the same
equation across all sky conditions, so that the two differ only through the
zenith-normalization term.  Distribution models formulated for radiance
usually reproduce measured sky luminance distributions with an accuracy
comparable to that of models developed specifically for luminance
\citep{Alshaibani:2020,Gracia:2011}.  Consequently, a model derived for the
radiance distribution can be taken to represent the corresponding luminance
distribution up to a scaling factor.  This is even more straightforward for
the present case than for daylight.  Unlike the daytime sky, where the
radiance observed in a given direction also receives contributions from light
scattered at very large distances, the relevant distances here are short.
}

\begin{figure*}[!ht]
 {
 \centering
 \leavevmode
 \includegraphics[width={0.45\linewidth}]{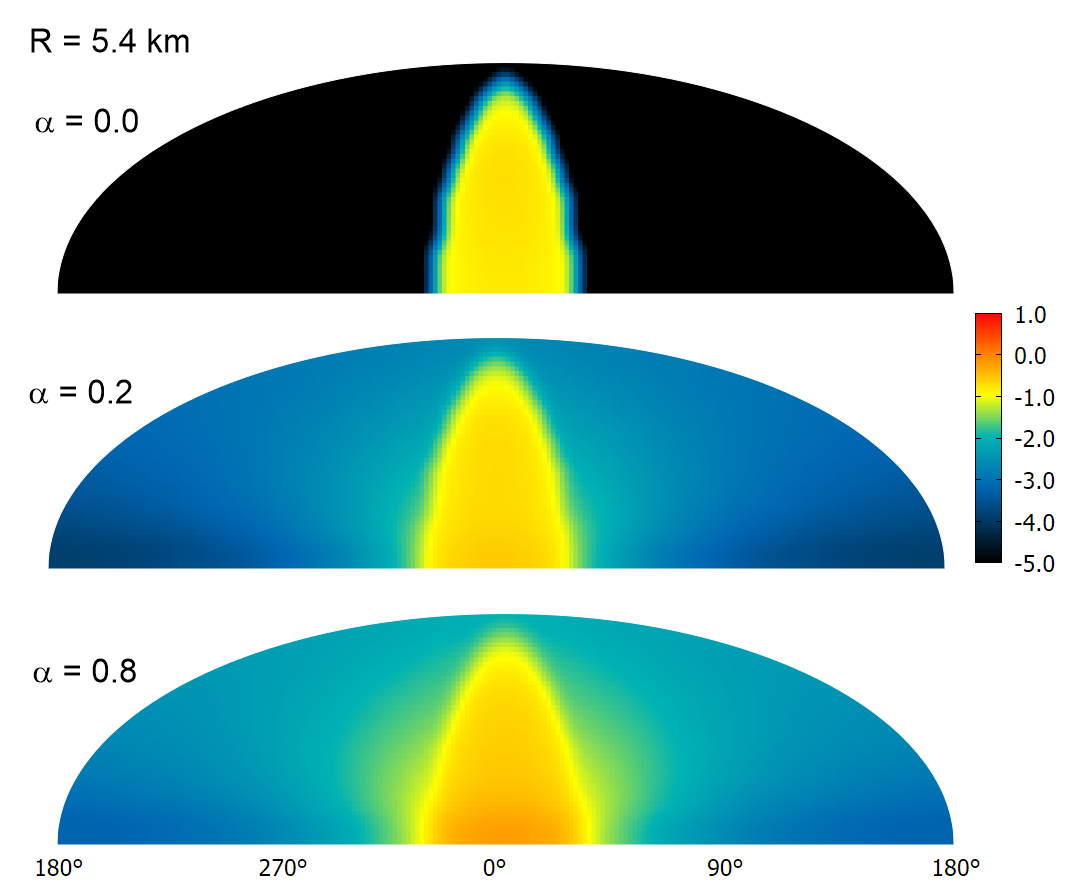}%
 \hfil
 \includegraphics[width={0.45\linewidth}]{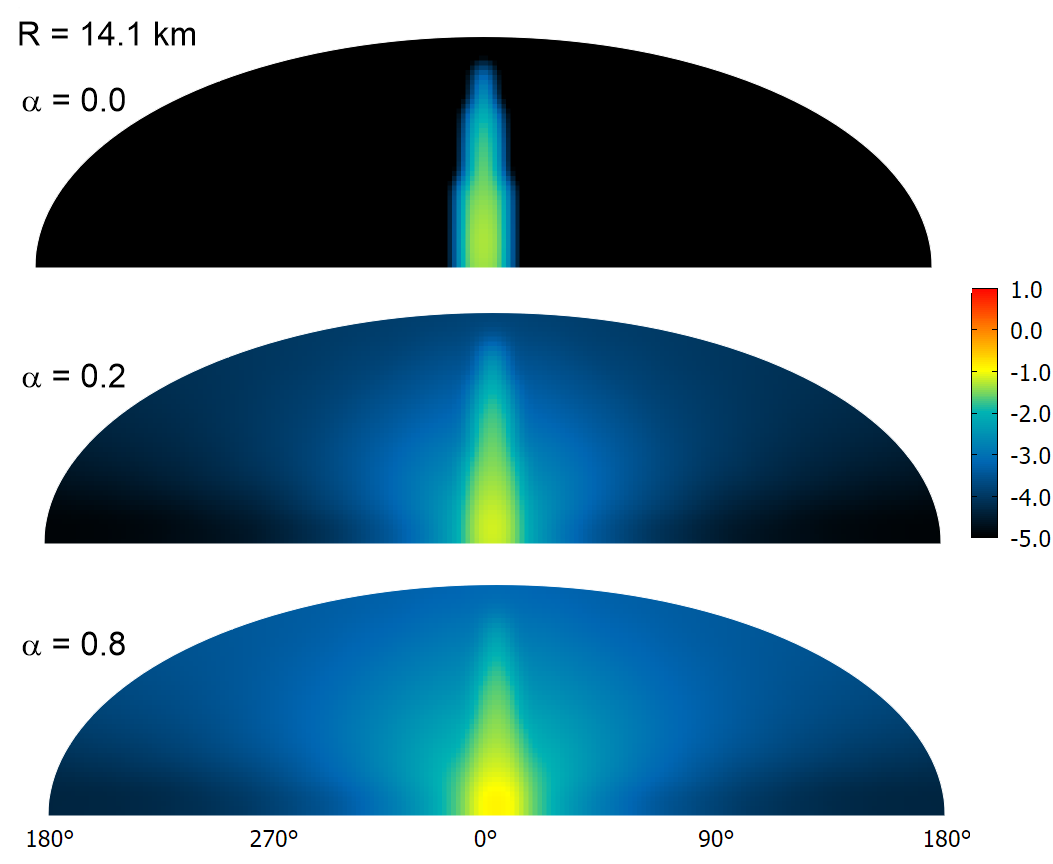}
 }
\caption{
Night-sky luminance maps in Hammer–Aitoff equal-area projection (in \cdmsq,
logarithmic color scale) for a perfectly absorbing surface ($\alpha = 0.0$),
a mean ground albedo ($\alpha = 0.2$), and a highly reflective
(snow-covered) surface ($\alpha = 0.8$), computed for aerosol optical depth
$\tau_{a} = 0.3$ and aerosol asymmetry parameter $g = 0.6$.
{\em Left:} an observer located at a horizontal distance of $R=$ 5.4\,km
from the axis of the light cone produced on the ground by a 54\,m mirror in
Earth orbit, shining straight down onto the surface.  The radius of the
directly illuminated area is 2.5\,km, so the observer is roughly 3\,km from
the edge of the lit portion of the ground.
{\em Right:} The same situation, but for an observer 14.1\,km from the
center of the illuminated circle on the Earth’s surface.
\label{fig:Fig_sky_g_0.6}
}
\end{figure*}

The computed night-sky luminance maps for observer distances of 5.4\,km and
14.1\,km from the beam center are shown in Fig.~\ref{fig:Fig_sky_g_0.6} in
Hammer--Aitoff equal-area projection, for three ground albedos ($\alpha =
0.0$, $0.2$, and $0.8$) and aerosol asymmetry parameter $g = 0.6$.  
\rev{
These
distances arise from the numerical configuration rather than from a
deliberate selection of particular observing scenarios.  The output nodes are
defined on an exponential (logarithmic) rather than a linear scale over the
modelling domain, which reflects the exponential character of the functions
governing radiative transfer, for which such sampling is natural.
}  For a
perfectly absorbing idealized surface ($\alpha = 0.0$), the sky luminance is
exclusively due to light scattering in the atmosphere, and decreases steeply
with increasing angular distance from the cone of light.  With no light
reflected from the surface, the contrast between the beam and its
surroundings is particularly pronounced.  The presence of ground-reflected
light ($\alpha > 0$) substantially modifies the luminance field.  The
Lambertian surface becomes a secondary source, redirecting diffuse radiation
into the entire upper hemisphere, resulting in a significant brightening of
the sky at the zenith and at intermediate elevation angles.  Near the
horizon the sky remains dark, consistent with the cosine law as the zenith
angle approaches $90\degr$.  To facilitate comparison and to document the
spatial decay of sky luminance with increasing observer distance ($R$), the
results for horizontal distances of $R=5.4$\,km
(Fig.~\ref{fig:Fig_sky_g_0.6}, left) and 14.1\,km
(Fig.~\ref{fig:Fig_sky_g_0.6}, right) are displayed on the same color scale. 
At 14.1\,km from the beam center, the luminance of the light cone near the
horizon can reach
approximately 0.15\,\cdmsq (14.7\,\msqarcsec) for a snow-covered surface,
while for a perfectly absorbing surface the corresponding value is
approximately half ($15.34\,\msqarcsec$). These are an order of magnitude brighter than
the sky would be due to the full moon, as we will show later in Sec.~\ref{sec:moon}.

Multiple scattering superimposes an additional, markedly more homogeneous
continuum upon the single-scattering luminance field, yet does not alter the
fundamental dominance of forward scattering in the overall angular
distribution of the scattered light.  When ground-reflected light is
present, photons travel along trajectories spanning all directions, so that
for any given viewing direction, there always exist photons undergoing
predominantly forward scattering.  This gives rise to a pronounced diffuse
background extending well beyond the geometric boundary of the light cone,
as clearly seen in Fig.~\ref{fig:Fig_sky_g_0.6} for $\alpha = 0.8$.

\begin{figure*}[!ht]
    \begin{minipage}[t]{0.5\textwidth}
        \includegraphics[width=\columnwidth]{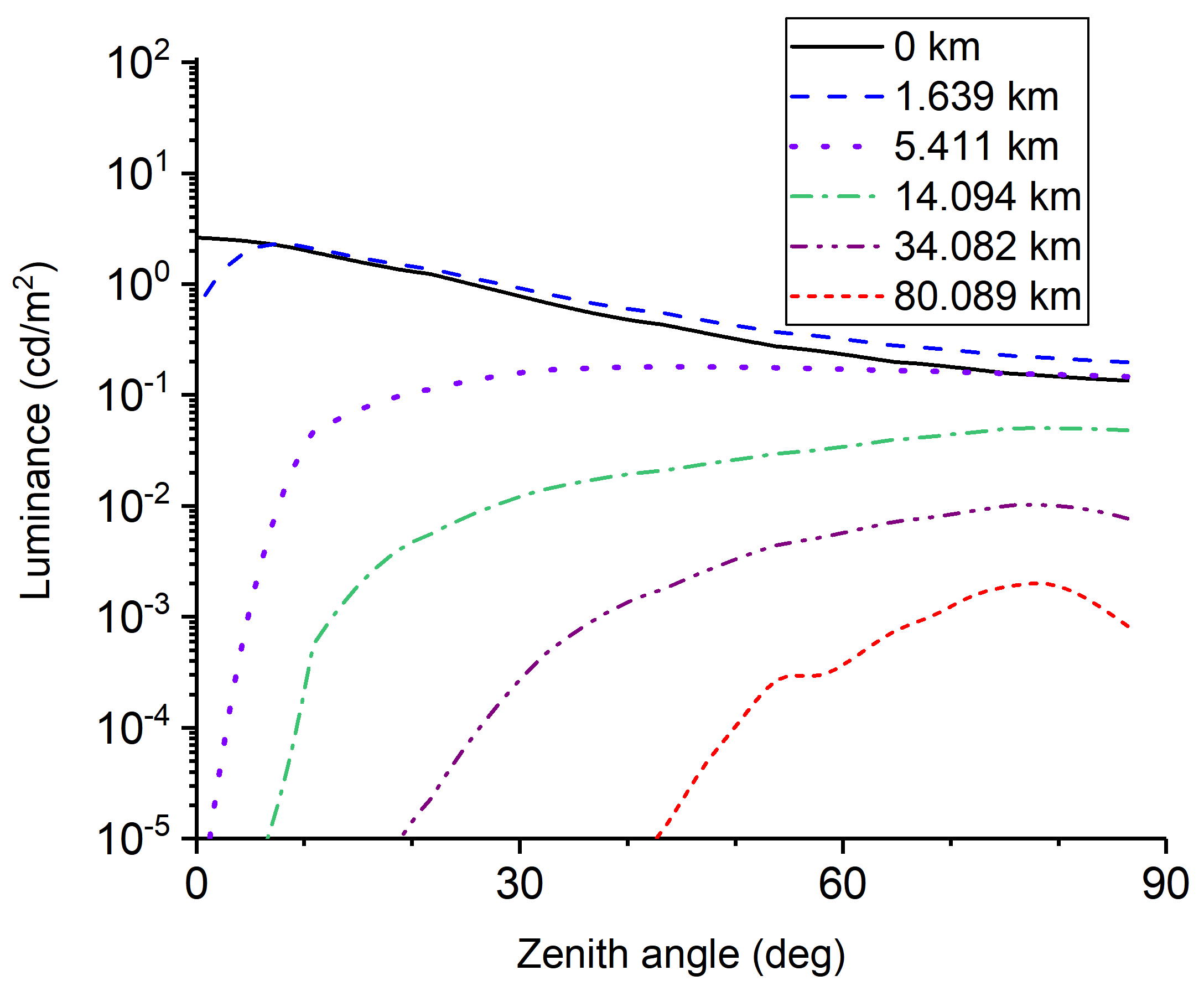}
    \end{minipage}%
    \begin{minipage}[t]{0.5\textwidth}
        \includegraphics[width=\columnwidth]{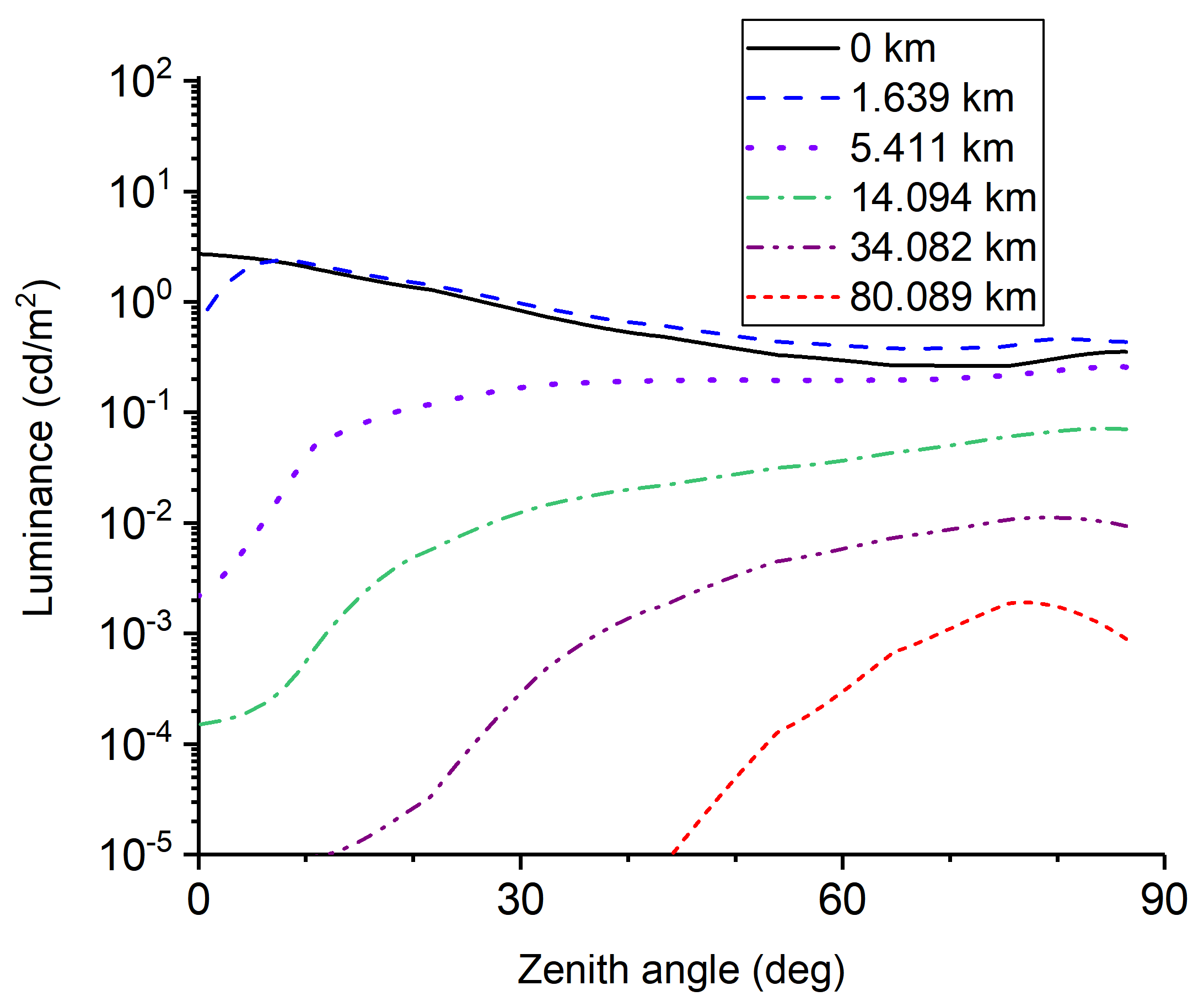}
    \end{minipage}
    \\[0.5ex]
    \begin{minipage}[t]{0.5\textwidth}
        \includegraphics[width=\columnwidth]{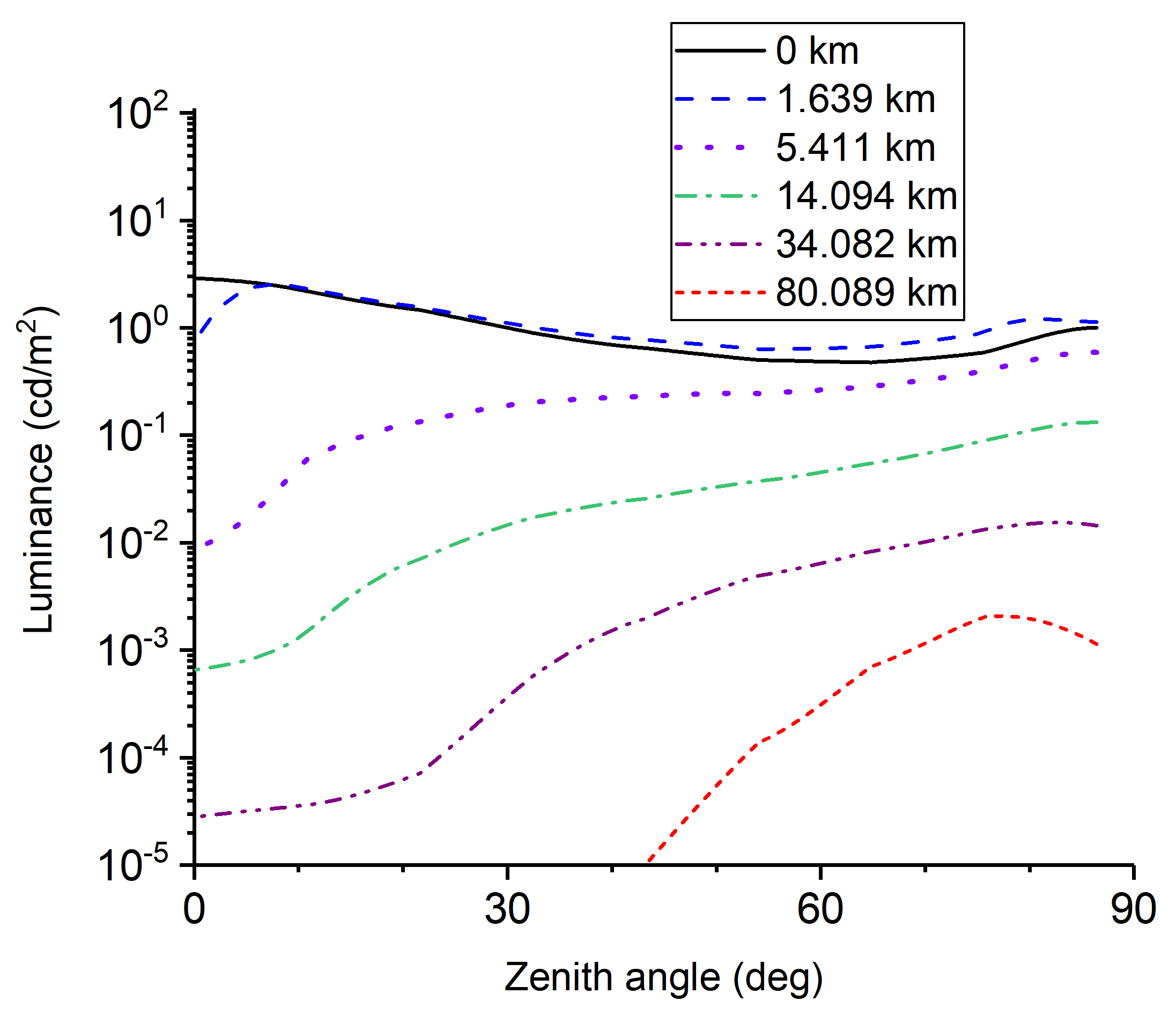}
    \end{minipage}%
    \begin{minipage}[t]{0.5\textwidth}
        \includegraphics[width=\columnwidth]{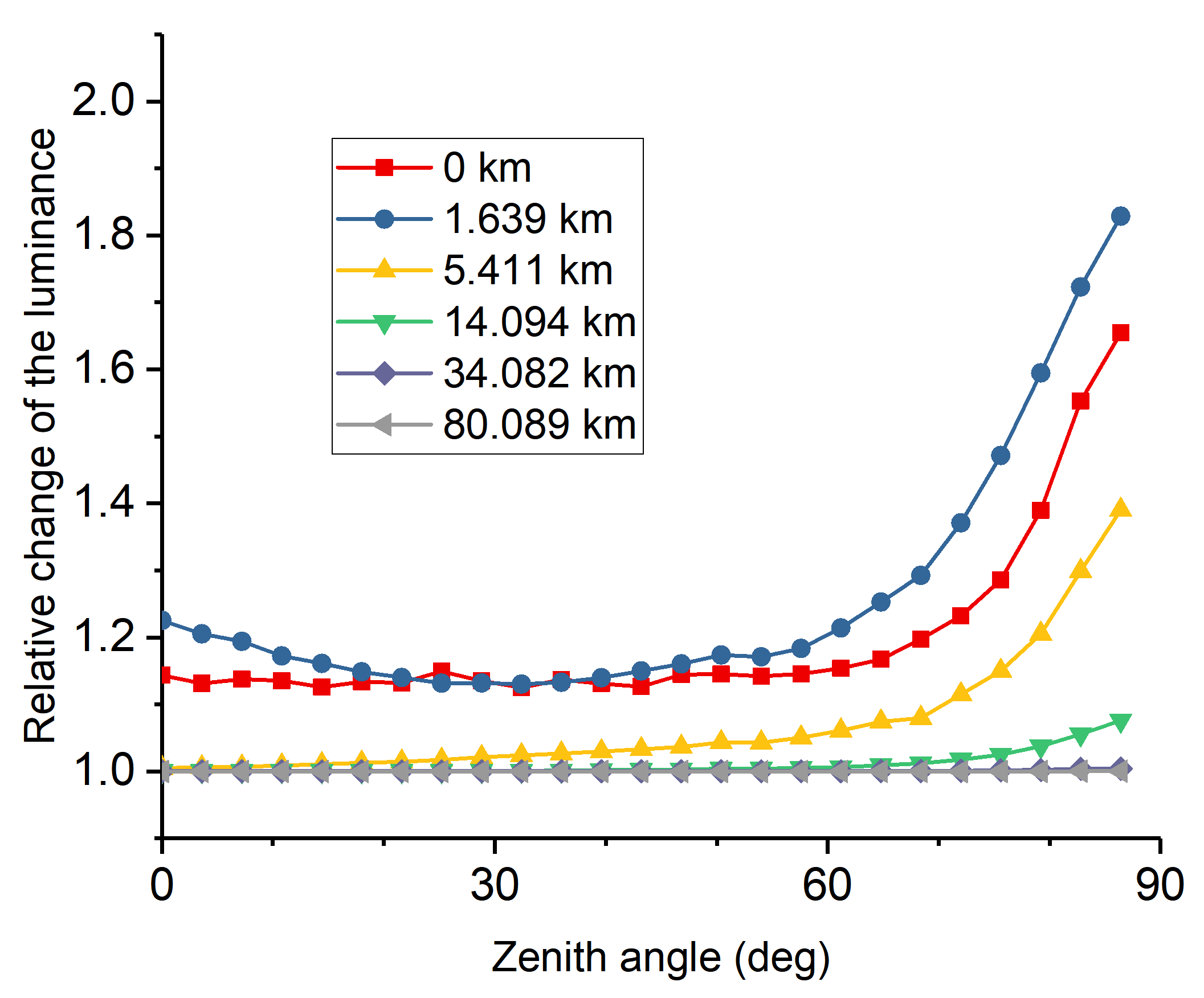}
    \end{minipage}
\caption{
{\em Top left:} The sky luminance (in \cdmsq) as a function of
zenith distance for an observer located at different distances from the
center of the beam produced by a 54\,m mirror at zenith. We assumed
$\alpha=0.0$ and $g=0.6$.
{\em Top right:} The same as top left but for $\alpha=0.2$.
{\em Bottom left:} The same as top left but for $\alpha=0.8$.
{\em Bottom right:} The factor of the luminance increase when transitioning from
$\alpha = 0.0$ to $\alpha = 0.2$.
\label{fig:sky_and_hor_g_0.6_A_0.2}
}
\end{figure*}

The influence of ground-reflected light manifests not only in the emergence
of a prominent luminance veil across the night sky, but also in an increase
of luminance of the light cone.  Although the contribution of reflected
light decreases steeply toward the horizon, multiply scattered light
modifies the light scene.  The contribution from higher scattering orders
increases with growing optical path \citep{vanDeHulst1980}.  While the
optical path for rays traveling vertically is short, the integral of the
atmospheric volume scattering coefficient for photons traversing the
atmosphere horizontally can be orders of magnitude larger.  The longer the
optical path, the more scattering events occur along the photon trajectory
from source to observer.  This effect is well illustrated by the bottom
right panel of Fig.~\ref{fig:sky_and_hor_g_0.6_A_0.2}, which shows the ratio
of sky luminance for $\alpha = 0.2$ to that for $\alpha = 0.0$.

Fig.~\ref{fig:sky_and_hor_g_0.6_A_0.2} also shows the sky luminance for
$\alpha = 0.0$ (top left panel), $\alpha = 0.2$ (top right panel), and
$\alpha = 0.8$ (bottom left panel) as a function of zenith angle, computed
for discrete observer distances from the center of the illuminated circular
patch of radius 2.5\,km.  Observers at $0$\,km and $1.6$\,km from the center
are directly within the artificially lit area, which is discussed later in
Sec.~\ref{sec:in}, while all remaining cases correspond to observers outside
the directly illuminated area.  Since the viewing direction is always toward
the beam axis, the optical path within the light cone for an observer at
$1.6$\,km exceeds that for an observer at the center.  The luminance is
computed for the diffuse component of downward radiation only, with the
direct beam contribution from the mirror excluded.

Therefore, the luminance profiles are primarily dictated by geometry, with a
stronger luminance enhancement occurring wherever the line of sight
intersects the axis of light cone.  For an observer outside the beam and for
$\alpha = 0.0$, the zenith luminance is determined solely by multiply
scattered light; first-order scattering contributes only when the viewing
direction crosses the beam.  The lower the altitude of this intersection,
the stronger the luminance enhancement, since both aerosol and molecular
concentrations decrease with height, reducing the scattering probability and
increasing the mean free path between successive scattering events.  The
presence of a reflected component manifests as an additional optical signal
over a relatively wide range of azimuths around the beam.

\section{Calculations within the beam}
\label{sec:in}

For an observer located within the directly illuminated region on the
Earth's surface, the lighting conditions differ significantly from those
outside the illuminated patch.  In addition to the direct illumination
estimated in Sec.~\ref{sec:backofenvelope}, the diffuse light field within
the circular patch is dominated by single-scattered radiation.  
\rev{
The
higher scattering orders contribute at most 1--4\% of the luminance 
and leave the angular radiance distribution unchanged.
}  
Light
scattering in the vertical atmospheric column is governed by the volume
scattering coefficient $k_\mathrm{sca}$ (unit m$^{-1}$), which depends on
altitude $h$ but is independent of scattering angle (see, e.g.,
\citealt{Ma:2021}, where it is denoted as $\beta$).  In a cloud-free
atmosphere, $k_{sca}$ is given by the sum of the aerosol component
$k_{sca,a}$ and the Rayleigh component $k_{sca,R}$.  Both components
decrease exponentially with altitude, with the rate of decrease
characterized by the molecular scale height ($H_{R}$) and the aerosol scale
height ($H_{a}$), respectively.  Since $H_{a}$ is typically several times
smaller than $H_{R}$, the vertical gradient of aerosol concentration is much
steeper than that of air density.  As a result, the lower troposphere plays
a key role in skyglow formation.  The integral of the Rayleigh scattering
coefficient over the vertical column from altitude $h$ to the top of the
atmosphere defines the Rayleigh optical depth $\tau_\mathrm{R}(h) =
\tau_\mathrm{R,0}\exp(-h/H_\mathrm{R})$, where $\tau_\mathrm{R,0}$ is the
Rayleigh optical depth at ground level.  Unlike molecules, which scatter
conservatively (i.e.~, the single scattering albedo is $\tilde{\omega}_{R}
\approx 1$), aerosols both scatter and absorb radiation, so that
$\tilde{\omega}_{a} < 1$ in general.  The aerosol volume scattering
coefficient can therefore be written as $k_\mathrm{sca,a} =
\tilde{\omega}_\mathrm{a}\,k_\mathrm{ext,a}$, where
$\tilde{\omega}_\mathrm{a}$ represents the fraction of energy scattered
relative to the total energy removed from the beam by aerosol particles.  By
analogy with the Rayleigh case, the aerosol optical depth is
$\tau_{a}(h)=\tau_{a,0}\exp(-h/H_{a})$.  The total optical depth at altitude
$h$ is then $\tau(h)=\tau_{R}(h)+\tau_{a}(h)$.

\begin{figure*}[!ht]
 {
 \centering
 \leavevmode
 \includegraphics[width={0.45\linewidth}]{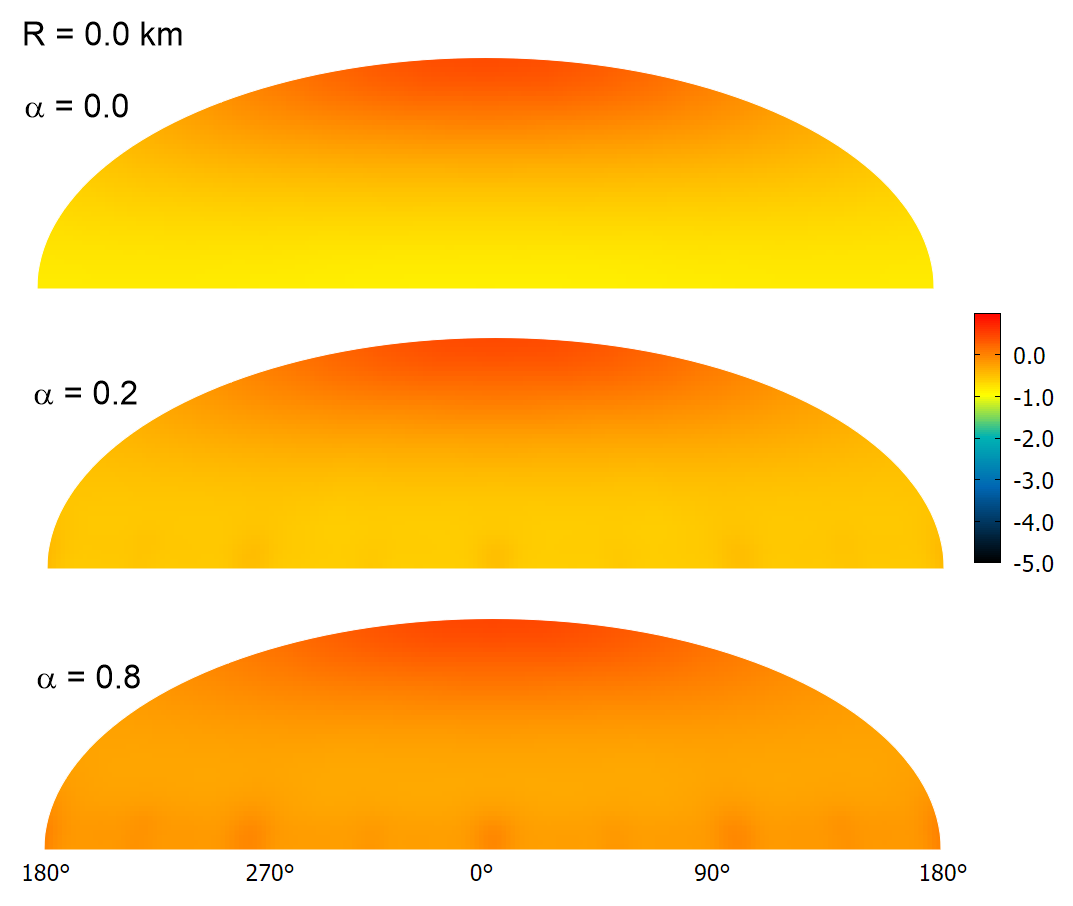}%
 \hfil
 \includegraphics[width={0.45\linewidth}]{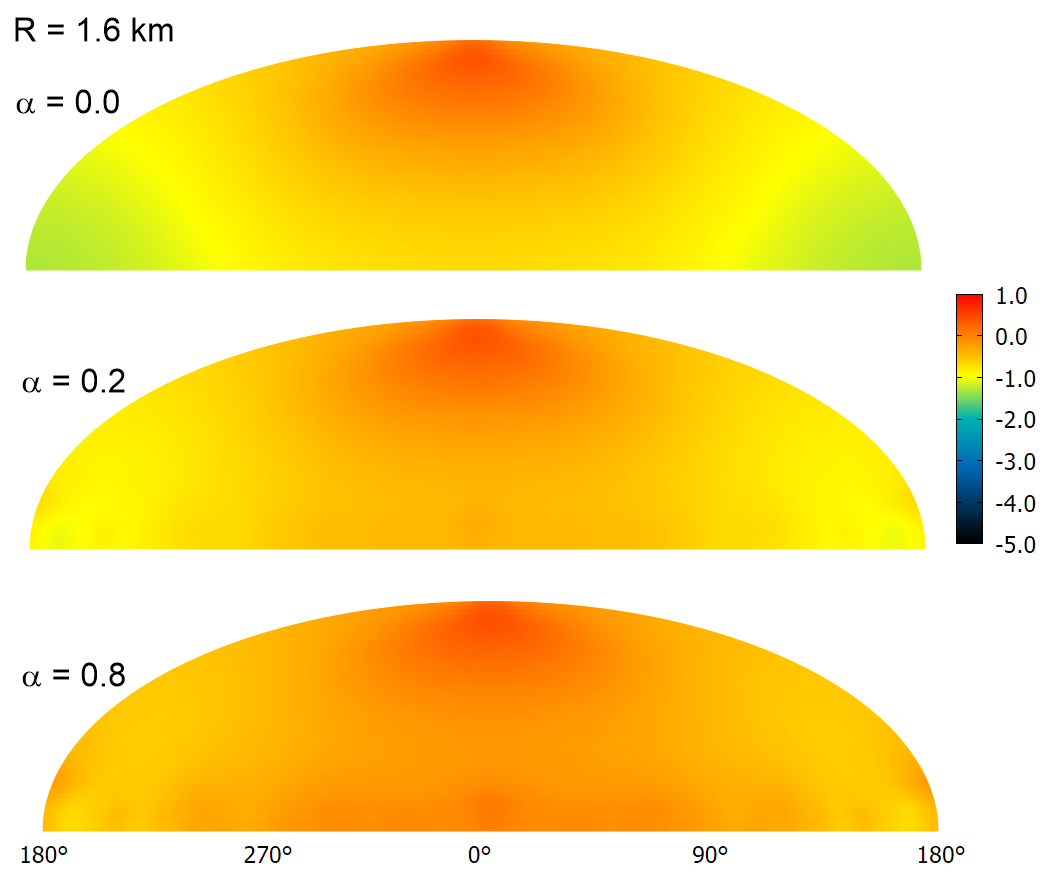}
 }
\caption{
The same as in Fig.~\ref{fig:Fig_sky_g_0.6}, but for an observer located
inside the illuminated area.
{\em Left:} an observer at the center of the illuminated region, directly
beneath the mirror.
{\em Right:} an observer 1.6\,km from the axis of the light cone.
\label{fig:Fig_lightcone_g_0.6}
}
\end{figure*}

The flux density of solar radiation at altitude $h$ is 
$$F(h)=F_\odot \frac{A_{proj}}{\pi \psi_{0}^{2}(H-h)^{2}}\exp\{-\tau(h)\},$$
where $\psi_{0}$ is the beam divergence ($\psi_{0} \ll 1\degr$).  At the
Earth's surface ($h=0$\,km), the denominator reduces to $\pi \psi_{0}^{2}
H^{2}$, which equals $A_{spot}$.  The amount of radiation scattered along a
path element $d\ell$ is
$$dI=\frac{P(\theta)}{4\pi}\,k_{sca}(h)\,F(h)\,d\ell,$$
where $P(\theta)$ is the phase function describing the angular distribution
of scattered radiation as a function of scattering angle $\theta$
\citep{Horvath:2014}.  In a vertically stratified atmosphere, it is
convenient to substitute $d\ell=dh/\cos z$, where $z$ is the zenith angle of
observation.  The scattering source term can be expanded as
$$ \frac{P(\theta)}{4\pi}k_{sca} = 
\frac{1}{4\pi}\left[k_{sca,R}(h)\,P_{R}(\theta)+
k_{sca,a}(h)\,P_{a}(\theta)\right].$$
In the Henyey--Greenstein formalism,
$$P_{a}(\theta)=\frac{(1-g^{2})}{(1+g^{2}-2g\cos\theta)^{3/2}}$$ \citep{Winkler:2022}. 
From the scattering geometry it follows that $\theta \approx z$, so the
Rayleigh phase function reduces to $P_{R}(z)=\frac{3}{4}(1+\cos^{2}z)$.  The
night-sky brightness at the observer's location is therefore 
$$I=\int
\exp\left\{-\frac{\tau_{a,0}}{\cos z} \left(1-e^{-\frac{h}{H_{a}}}\right) -\\ 
\frac{\tau_{R,0}}{\cos z}\left(1-e^{-\frac{h}{H_{R}}}\right)\right\}dI,$$ where $\tau_{a,0}$ and
$\tau_{R,0}$ are the aerosol and Rayleigh optical depths at the ground,
respectively.  The integration extends from the Earth's surface to the TOA.
The computational results are demonstrated as luminance maps in
Fig.~\ref{fig:Fig_lightcone_g_0.6} for an observer located directly beneath
the mirror (left panel) and for an observer at approximately $1.6$\,km from
the beam axis (right panel).  

As follows from the 3D geometry, the sky luminance distribution for an
observer at the center of the illuminated patch ($R = 0$\,km) is azimuthally
symmetric (Fig.~\ref{fig:Fig_lightcone_g_0.6}, left).  With increasing
ground albedo ($\alpha$), the diffuse background at low elevation angles
brightens noticeably.  Near zenith, however, the sky luminance is much less
sensitive to the ground albedo, altogether very bright, but with a small
range: 2.6\,\cdmsq\ ($11.56$\,\msqarcsec, for $\alpha = 0.0$) to
2.9\,\cdmsq\ ($11.44$\,\msqarcsec\ for $\alpha = 0.8$).

For an observer displaced from the beam axis ($R = 1.6$\,km), the sky
brightness in zenith does not noticeably change.  However, the luminance
distribution becomes azimuthally asymmetric
(Fig.~\ref{fig:Fig_lightcone_g_0.6}, right).  Assuming $\alpha = 0.0$, the
near-horizon luminance in the direction toward the beam axis (i.e.  for
azimuth angle $A=0\degr$) reaches approximately 0.2\,\cdmsq, while in the
opposite direction ($A=180\degr$) it falls to below 0.05\,\cdmsq.  This
luminance disproportion broadly correlates with the ratio of the mean
optical path lengths traversed by the beam in the respective viewing
directions.  The azimuthal asymmetry is somewhat reduced at higher albedo
values such as for fresh snow cover ($\alpha = 0.8$), though it remains
clearly present.  However, in the latter case the near-horizon luminance
reaches approximately 1.4\,\cdmsq\ toward the beam axis and
$0.2$--$0.4$\,\cdmsq\ in the opposite direction.

\section{Comparison to the illumination by the full moon}
\label{sec:moon}

The lighting conditions of a night-time atmosphere illuminated by the Moon
were modeled using the UniSky Simulator \citep{Kocifaj:2015}, originally
developed for daylight studies.  This radiative transfer tool is based on
the modified theory of successive orders of scattering \citep{Kocifaj:2012}. 
The Moon is treated as a source of parallel rays, analogous to the Sun,
given its distance from the Earth.  This guarantees spatial illuminance
homogeneity, so that the boundary condition for the incident radiation field
at any point on the TOA can be written as $\delta(z - z_0)\,\delta(A -
A_0)\,F_\odot$, where $z_0$ and $A_0$ are the zenith and azimuth angles of
the Moon, and $F_\odot$ is the net flux density of moonlight.  The radiative
transfer problem is then solved subject to boundary conditions defined by
this external source at the TOA and a Lambertian reflecting surface at the
lower boundary.

To compare the diffuse light field produced by the mirror with that of the
Moon, we performed the calculations with the following parameters: ground
albedo $\alpha = 0$, $0.2$, and $0.8$, aerosol asymmetry parameter $g = 0.6$
and $g = 0.8$, single-scattering albedo of aerosol particles $= 0.9$,
aerosol optical depth at the reference wavelength of $500$\,nm $= 0.3$,
aerosol scale height $= 2$\,km, and scale height of the molecular atmosphere
$= 8$\,km.  The results for moonlit sky include first- and second-order
scattered radiation, including the contribution from ground-reflected light.

\begin{figure*}[!ht]
 {
 \centering
 \leavevmode
 \includegraphics[width={0.31\linewidth}]{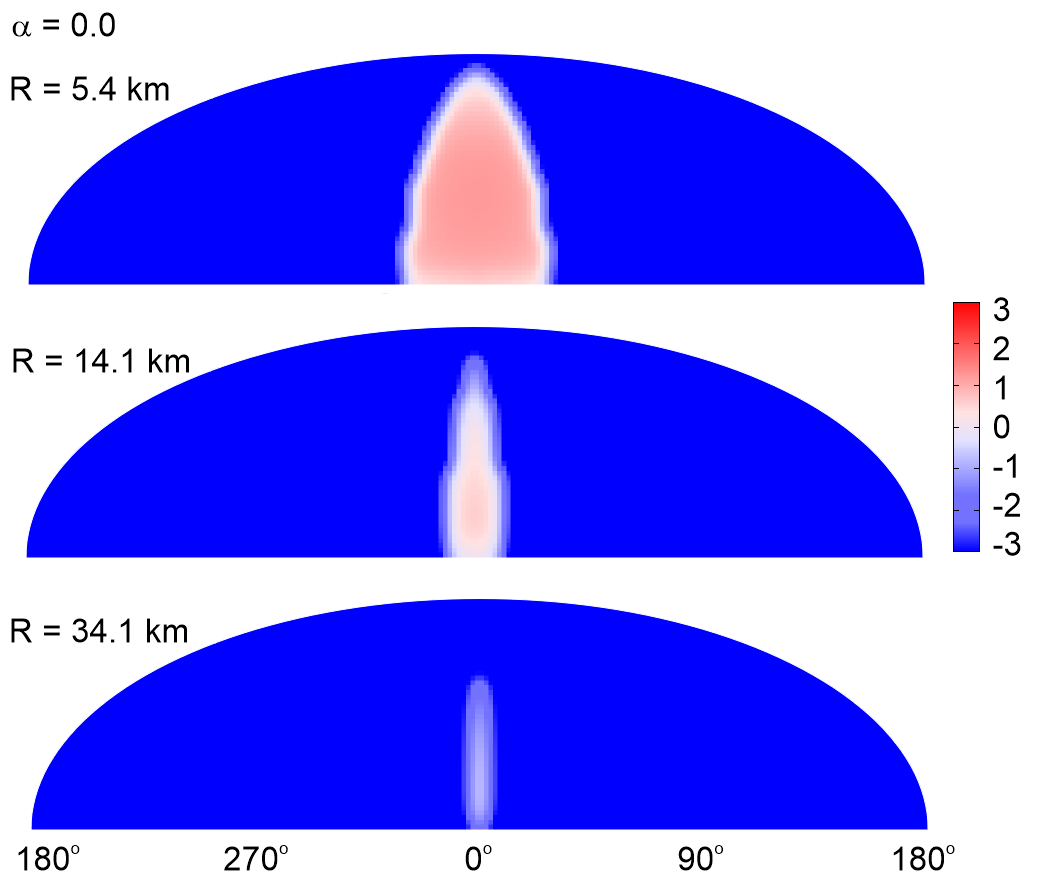}%
 \hfil
  \includegraphics[width=0.31\linewidth]{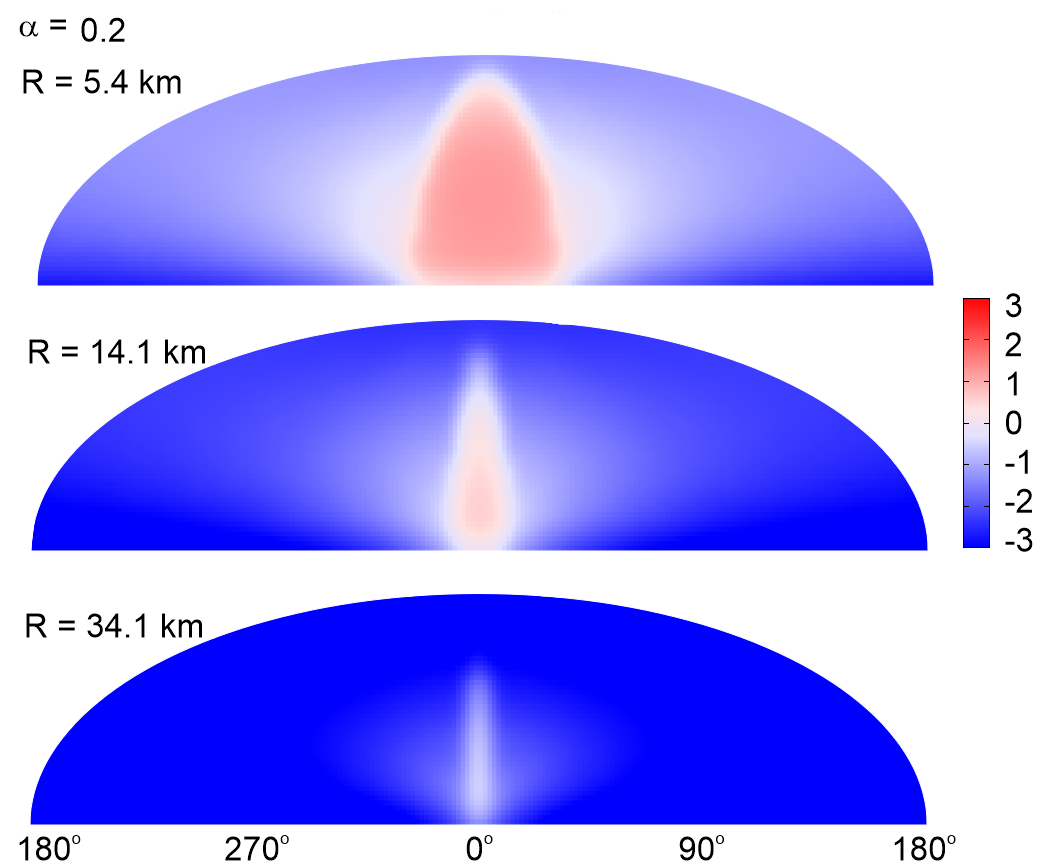}%
  \hfil
 \includegraphics[width={0.31\linewidth}]{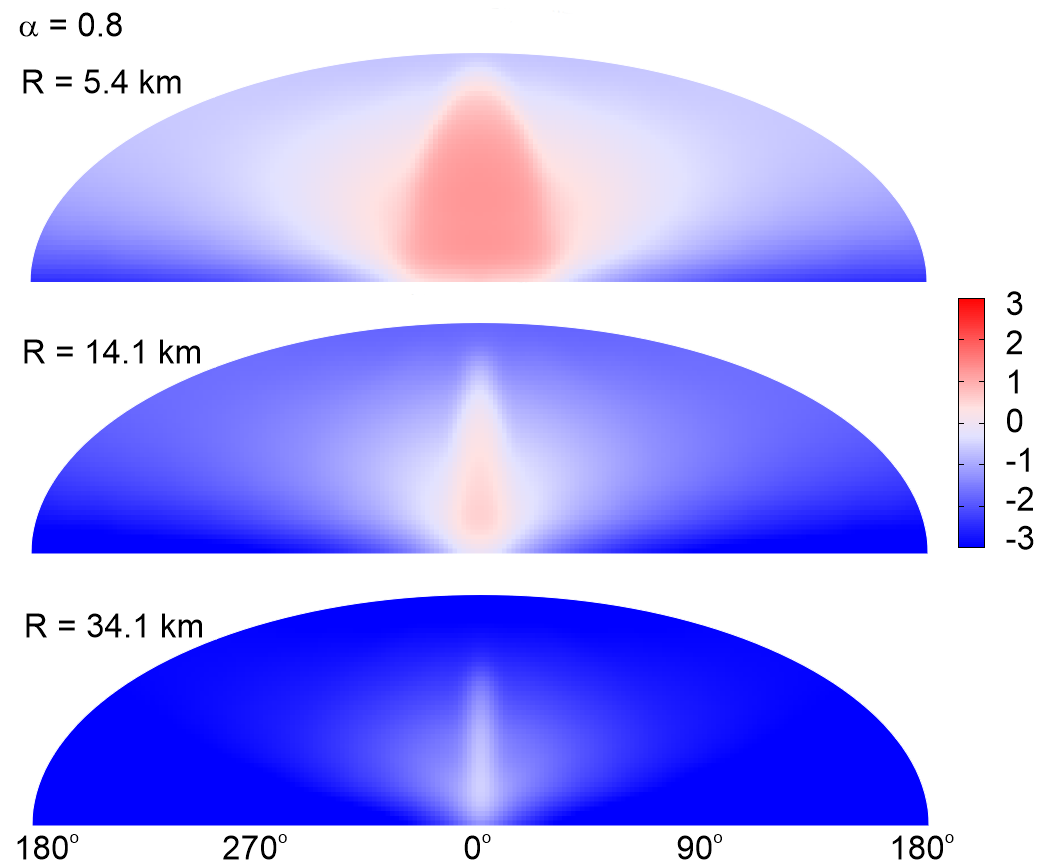}
 }
 \caption{
The same as in Fig.~\ref{fig:Fig_sky_g_0.6}, but showing the ratio of the
diffuse sky luminance produced by the mirror to that of the full-Moon-lit
sky (blue: mirror $<$ Moon; red: mirror $>$ Moon; white: mirror $\approx$
Moon).  
{\em Left:} perfectly absorbing surface ($\alpha = 0.0$);
{\em Middle:} Typical ground albedo of ($\alpha = 0.2$); 
{\em Right:} highly reflective snow-covered surface ($\alpha = 0.8$).  From
top to bottom, the observer is located at horizontal distances of 5.4\,km,
14.1\,km, and 34.1\,km from the axis of the light cone, all outside the
directly illuminated area (radius 2.5\,km).
\label{fig:Fig_relative2monnlight_g_0.6}
}
\end{figure*}

We then derived the ratio of the sky brightness due to the mirror and the
Moon (both at zenith) to perform a quantitative comparison.  As we showed in
Sec.~4, for an observer within the beam, the sky brightness ranges between
0.1\,\cdmsq\ (15\,\msqarcsec) and 2.9\,\cdmsq\ (11.4\,\msqarcsec), which
exceeds the luminance of the moonlit sky (0.01\,\cdmsq\ or 18\,\msqarcsec) by
factors of 10 to 300.  Thus, our analysis pertains to observers outside of
the beam.  To distinguish the results from the absolute luminance maps
presented in Figs.~\ref{fig:Fig_sky_g_0.6}--\ref{fig:Fig_lightcone_g_0.6},
the ratio of diffuse sky luminances produced by the mirror and the Moon
(both at zenith) is shown on a different color scale in
Fig.~\ref{fig:Fig_relative2monnlight_g_0.6}, where blue represents cases in
which the diffuse sky luminance due to the mirror is negligible compared to
that of the moonlit sky, and red represents the opposite extreme.  Cases
where the two luminances are comparable are shown in white.  Results are
presented for the three albedo values considered in this work, i.e., for a
perfectly absorbing surface (Fig.~\ref{fig:Fig_relative2monnlight_g_0.6},
left), a typical ground albedo (Fig.~\ref{fig:Fig_relative2monnlight_g_0.6},
middle), and a highly reflective snow-covered surface
(Fig.~\ref{fig:Fig_relative2monnlight_g_0.6}, right).
In general, the luminance of the artificial light from the 54\,m mirror with
a ground albedo of $\alpha=0.2$ far exceeds the luminance from the diffuse
moonlit sky for most of the sky, except the horizon opposite to the beam,
even at a distance of 14.1\,km.  This dominance is even more pronounced for
higher albedos ($\alpha=0.8$).
The luminance of the vertical beam of light against the diffuse moonlit sky
background decreases with increasing distance from the center of the
mirror-illuminated patch.  At a distance of 5.4\,km, the beam luminance
exceeds that of the moonlit sky by a factor of $\sim 15$ in the beam
direction; at 14.1\,km this ratio drops to $\sim 4$; and at 34.1\,km the
beam luminance falls to roughly half the moonlit sky background.  These
ratios vary only slightly (less than 10\%) with ground albedo at distances
below $\sim 20-30$\,km.  However, at distances exceeding $\sim 30$\,km, the
beam-to-moonlight luminance ratio for a snow-covered surface can be up to
40\% higher than for an ideally dark surface.  

An elevated ground albedo causes the surrounding landscape to reflect a
fraction of moonlight back into the atmosphere.  On the other hand, the
mirror in its orbit illuminates only a small patch of the Earth's surface. 
As the observer distance from the beam increases, atmospheric extinction
causes a more pronounced decay of beam intensity, while the higher diffuse
background from ground-reflected moonlight lowers the luminance ratio and
thus reduces the contrast of the light cone against the moonlit sky.

Lambertian reflection further disadvantages photons traversing the
atmosphere along highly inclined trajectories, while the contribution 
of multiply scattered moonlight near the horizon produces a continuous
luminance background. As a result, the ratio of the mirror-originated
to moonlit sky luminance shows a minimum near the horizon, which
gradually gives way to a broad luminance veil in the sky quadrant
containing the beam. This luminance veil has an azimuthal
width ranging from a few tens of degrees at a distance of 
$R \sim 30$\,km to more than $100\degr$ at the distance 
of $\sim 5$\,km. The luminance within this veil receives comparable
contributions from scattered moonlight and scattered beam light.

\section{Conclusions}
\label{sec:conc}

In this paper, we calculated the sky brightness due to a single,
$54\times54$\,m reflective space mirror, proposed by ReflectOrbital, as a
long-term solution for illuminating $\sim$2.5\,km radius spots on Earth. 
The satellite would appear as a -16.7 magnitude point-like source, i.e.,
about 4 magnitudes (40 times) brighter than the full moon, illuminating the
ground at $\sim9.5$\,lux (street-light level).  The brightness of the night
sky will be increased due to Rayleigh and aerosol scattering of both the
incoming beam and the light scattered back from the ground.  Even without
detailed calculations, by scaling from everyday experience, the resulting
light pollution should be enormous due to 2.5\,km spot evenly illuminated at
street-light level, far exceeding the uplight of a town with a similar surface
area (given that street lighting is highly localized and streets only occupy
$\sim10$\% surface area).  This is confirmed by our detailed calculations, which
we carry out for observers both outside and inside the beam.

For observers outside the beam we used the Modified Successive Orders of
Scattering (MSOS) model from \citep{kocifaj:2018,kocifaj:2023} to compute
the field of scattered radiation at any point in 3D space.  We carried out
these calculations for three different surface albedos ($\alpha=0, 0.2,
0.8$), and two scattering asymmetry parameters ($g=0.6, 0.8$).  We computed
night-sky luminance maps for different observer distances, ranging from
5.4\,km to 80\,km from the center of the beam.  
We show that the sky brightness for average ground reflection ($\alpha=0.2$)
will be much higher in almost all of the sky, except the horizon opposite to
the beam, for observers located at 14\,km from the center of the beam,
meaning that the sky will be more light-polluted thandue to the full
moonMoon in a circle with an approximate radius of 14\,km (typical albedo) to
30\,km (snow-covered terrain).
The sky will be brighter than the full-moon lit sky at 34\,kms
from the center of the beam in the direction of the beam, appearing as a
brightly lit pillar with \rev{large} side-lobes ($\pm45\arcdeg$ in azimuth and
up to zenith).

For observers in the beam the diffuse light field is dominated by
single-scattered radiation.  We took into account scattering from both
aerosols and atmospheric molecules.  The sky brightness (up to 2.9\,\cdmsq\
or 11.45\,\msqarcsec) will typically far exceed that of the full-moon lit sky
down to very close to the horizon (\tabr{skybr}).  The level of sky brightness will be
similar to the dusk sky in zenith shortly after sunset, when the sun is at
$\sim-5\arcdeg$ elevation \citep{Patat:2006}.  Not even the brightest stars
remain visible.

\begin{table}
    \centering
    \begin{tabular}{ccc}
    R        & Lum ($\alpha = 0.2$)   & Lum ($\alpha = 0.8$)\\
    (km)     & (\cdmsq) (\msqarcsec) & (\cdmsq) (\msqarcsec)\\\hline\hline
    0        &  2.71   (11.52) &  2.89 (11.45) \\
    1.6      &  2.62   (11.55) &  2.77 (11.49) \\
    5.4      &  0.25   (14.11) &  0.59 (13.17) \\
    14.1     &  0.07   (15.49) &  0.13 (14.82) \\
    34.1     &  0.011  (17.50) &  0.015 (17.16)
    \end{tabular}
    \caption{The maximum sky luminance in \cdmsq\ for selected scenarios of 
    different observer distances from the center of the beam from the 54\,m mirror. 
    These calculations assume $g = 0.6$. For comparison, the median sky brightness 
    due to the full moon is 0.014\,\cdmsq\ ($\alpha=0.2$) and 0.02\,\cdmsq\ ($\alpha=0.8$).}
    \label{tab:skybr}
\end{table}

Altogether, within the illuminated patch, the night sky is lost.  Outside
the beam, the night sky is seriously compromised to distances up to 30\,kms,
with the light pillar of the beam and the ground-reflection of light scatter
visible from even larger distances.  Our calculations present a best-case
scenario by assuming a completely cloud-free atmosphere.  With a cirrus
cloud-deck at e.g.~10\,km altitude illuminated by the beam, the light
pollution \rev{could} be visible from much larger distances.

\rev{
The influence of clouds on night-sky brightness is complex.  For an
atmosphere illuminated from below by ground-based sources we have shown that
this influence depends on the emission function of the source and on its
distance from the observer.  We have also shown that a transition zone
exists, in which the amplifying effect of clouds turns into a weakening one
\citep{Kocifaj:2025pnas,Kocifaj:2025mnrasl}.  However, for a light source above the
atmosphere, such as from a satellite orbiting the Earth, 
the model would be substantially different and would require separate
development.
}

\rev{
Dense clouds, which for ground-based sources strongly reshape the
night-sky brightness distribution, would here act as a barrier and would
usually block the light from reaching the surface.  Thin (cirrus)
clouds would alter the radiation field computed for a clear sky in two ways,
removing part of the radiation and scattering another part into other
directions.  The outcome would depend strongly on the actual
source-cloud-observer geometry and on the shape and opacity of the cloud in
its different parts.  Such conditions can therefore be captured only through
a statistical approach, describing the mean sky state expected for a given
cloud type, sky coverage fraction, cloud height, vertical distribution, and
opacity. These calculations are beyond the scope of the current paper. 
}

\section*{Acknowledgements}
\begin{acknowledgements}

This work was supported by the Slovak Research and Development Agency (grant
number APVV-22-0020) and by the Scientific Grant Agency VEGA (grant number
2/0009/24).  This work was also supported by the Office of Sustainability at
Princeton University.  The authors thank Dr.~Joel Hartman for his careful
reading of the manuscript and his detailed feedback. GÁB wishes to thank
Dr.~Emilio Falco and Sarah Thiele for scientific discussions on the topic.

\end{acknowledgements}

\software{Modified Successive Orders of Scattering (MSOS) model
\citep{kocifaj:2018,kocifaj:2023} and UniSky Simulator
\citep{Kocifaj:2012,Kocifaj:2015}}


\bibliography{robib}
\bibliographystyle{aasjournal}

\end{document}